\documentstyle[epsfig]{aipproc}

\begin{document}
\title{Partonic structure of $\gamma_L^*$ in hard collisions}

\author{Ji\v{r}\'{\i} Ch\'{y}la$^*$}
\address{$^*$
Research Center for Particle Physics, Institute of Physics
of the Academy of Sciences\\
18221 Na Slovance 2, Prague 8, Czech Republic}
%
\maketitle
\begin{abstract}
Manifestation of QCD improved partonic structure of longitudinally
polarized virtual photons in hard collisions is discussed. As an
example, dijet production in ep collisions at HERA is investigated
in detail.
\end{abstract}

\section*{Introduction}
\label{intro}
In this talk I discuss phenomenological consequences of
QCD improved partonic structure of longitudinally polarized virtual
photons ($\gamma_L^*$), concentrating on LO QCD calculations of dijet
production in ep collisions at HERA. Some of the results presented
here are discussed in detail in \cite{long,plb,friberg2}. The role
of resolved $\gamma_L^*$ in NLO QCD calculations will be addressed
elsewhere \cite{epjc}. I start by recalling the virtue of extending
the concept of partonic ``structure'' to virtual photons
\cite{prd,friberg}:
\begin{itemize}
\item In principle, the concept of partonic structure of
virtual photons can be dispensed with as higher order QCD corrections
to cross sections of processes involving virtual photons in the initial
state are well--defined and finite even for massless partons.
\item In practice, however, the concept of {\em resolved virtual
photon} is extraordinarily useful as
it allows us to include the resummation of higher order QCD effects
that come from physically well--understood region of (almost)
parallel emission of partons off the quarks and antiquarks coming
from the primary $\gamma^*\rightarrow q\overline{q}$ splitting
and subsequently participating in hard processes.
\end{itemize}
For the virtual photon, as opposed to the real one, its parton
distribution functions (PDF) can
therefore be regarded as ``merely'' describing higher order
perturbative effects and not their ``genuine'' structure. Although
this distinction between the content of PDF of real and virtual photons
exists, it does not affect the {\em phenomenological}
usefulness of PDF of the virtual photon. As shown in \cite{prd}
the nontrivial part of the contributions of resolved $\gamma_T^*$ to
NLO calculations of dijet production at HERA is large and affects
significantly the conclusions of phenomenological analyses of
existing experimental data. Taking into account resolved $\gamma_L^*$
turns out to be phenomenologically important as well.

\section*{Parton distribution functions of $\gamma_L^*$ in QCD}
\label{sec:qedqcd}
Most of the present knowledge of the structure of the photon
comes from experiments at ep and e$^+$e$^-$ colliders, where
the incoming leptons act as sources of transverse ($\gamma_T^*$)
and longitudinal ($\gamma_L^*$) virtual photons of virtuality $P^2$
and momentum fraction $y$. To order $\alpha$ their respective
unintegrated fluxes are given as
\begin{eqnarray}
f^{\gamma_T^*}(y,P^2) & = & \frac{\alpha}{2\pi}
\left(\frac{1+(1-y)^2)}{y}\frac{1}{P^2}-\frac{2m_{\mathrm e}
^2 y}{P^4}\right),
\label{fluxT} \\
f^{\gamma_L^*}(y,P^2) & = & \frac{\alpha}{2\pi}
\frac{2(1-y)}{y}\frac{1}{P^2}.
\label{fluxL}
\end{eqnarray}
Phenomenological analyses of interactions of virtual photons
and their PDF have so far concentrated on $\gamma_T^*$.
Neglecting longitudinal photons is
a good approximation for $y\rightarrow 1$, where
$f^{\gamma_L^*}(y,P^2)\rightarrow 0$, as well as for small
virtualities $P^2$, where PDF of $\gamma_L^*$ vanish by gauge
invariance. But how small is ``small'' in fact? For instance,
should we take into account the contribution of $\gamma^*_{L}$ to
jet cross sections in the region $E_T\gtrsim 5$
GeV, $P^2\gtrsim 1$ GeV$^2$, where most of the data on virtual
photons obtained in ep collisions at HERA come from? The present
paper is devoted primarily to addressing this question.

In pure QED and to order $\alpha$ the probability of finding inside
$\gamma_L^*$ of virtuality $P^2$ a quark with mass $m_q$, charge $e_q$,
momentum fraction $x$ and virtuality $\tau\le M^2$, is given, in units
of $3e_q^2\alpha/2\pi$, as \cite{prd}
\begin{equation}
q^{\mathrm {QED}}_L(x,m_q^2,P^2,M^2)=
\frac{4x^2(1-x)P^2}{\tau^{\mathrm {min}}}
\left(1-\frac{\tau^{\mathrm {min}}}{M^2}\right),
\label{fullresult}
\end{equation}
where $\tau^{\mathrm {min}}=xP^2+m_q^2/(1-x)$.
The quantity defined in (\ref{fullresult}) has a clear physical
interpretation: it describes the flux of quarks that are almost
collinear with the incoming photon and ``live'' longer than $1/M$.
For $\tau^{\mathrm {min}}\ll M^2$ (\ref{fullresult}) simplifies to
\begin{displaymath}
q^{\mathrm {QED}}_L(x,m_q^2,P^2,M^2)=
\frac{4x^2(1-x)P^2}{xP^2+m_q^2/(1-x)},
\end{displaymath}
which for $x(1-x)P^2\gg m_q^2$ further reduces to
\begin{equation}
q^{\mathrm {QED}}_L(x,0,P^2,M^2)=4x(1-x),
\label{virtualL}
\end{equation}
whereas for $x(1-x)P^2\ll m_q^2$
\begin{displaymath}
q^{\mathrm {QED}}_L(x,m_q^2,P^2,M^2)\rightarrow
\frac{P^2}{m_q^2}4x^2(1-x)^2.
\end{displaymath}
QCD corrections to QED expressions for PDF of $\gamma_L^*$ have been
derived in leading-logarithmic approximation in the region
$1\lesssim P^2\ll M^2$ in \cite{plb}.
By ``leading--log'' I mean resummation of the terms
$(\alpha_s\ln M^2)^k$ at each order $k$ of perturbative QCD. Note
that for $\gamma_T^*$ there is one power of $\ln M^2$ more at each
order of $\alpha_s$, the additional one coming from the primary
QED $\gamma^*\rightarrow q\overline{q}$ splitting. In the case of
$\gamma_L^*$ the analogous splitting gives rise to ``constant''
term (\ref{virtualL}), hence the absence of this log.
The resulting expressions exhibit typical hadronic form of scale
dependence and contain $\Lambda_{\mathrm{QCD}}$ as the only free
parameter. QCD effects thus suppress quark distribution functions
$q_L^{\mathrm{QED}}(x,P^2,M^2)$ at large $x$ and enhance it on the
other hand for $x\lesssim 0.4$. Moreover, they generate sizable
gluon distribution function, absent in QED. The presence of the
term proportional to $\ln M^2$ in the expressions for $q_T$ in
both QED and QCD implies the dominance of $\gamma_T^*$ at large $M^2$,
but one would have to go to very large $M^2$ for $\gamma_L^*$ to become
really negligible with respect to $\gamma_T^*$. For fixed $M^2$ the
relative importance of $\gamma_L^*$ with respect to $\gamma_T^*$ grows
with $P^2$, but to retain clear physical meaning of PDF we stay
throughout this paper in the region where $P^2\ll M^2$. The lower
bound $1~{\mathrm{GeV}}^2\lesssim P^2$ ensures that hadronic parts
of QCD improved PDF of $\gamma_L^*$, which have not been taken into
account in \cite{plb}, can be safely neglected.

\section*{$\gamma_L^*$ in hard collisions}
The relevance of resolved $\gamma_L^*$ in hard collisions of virtual
photons depends on the theoretical framework we are working in.
In this talk I will stay within the framework of LO QCD calculations of
dijet production at HERA.
\begin{figure}\unitlength 1mm
\begin{picture}(160,70)
\put(0,0){\epsfig{file=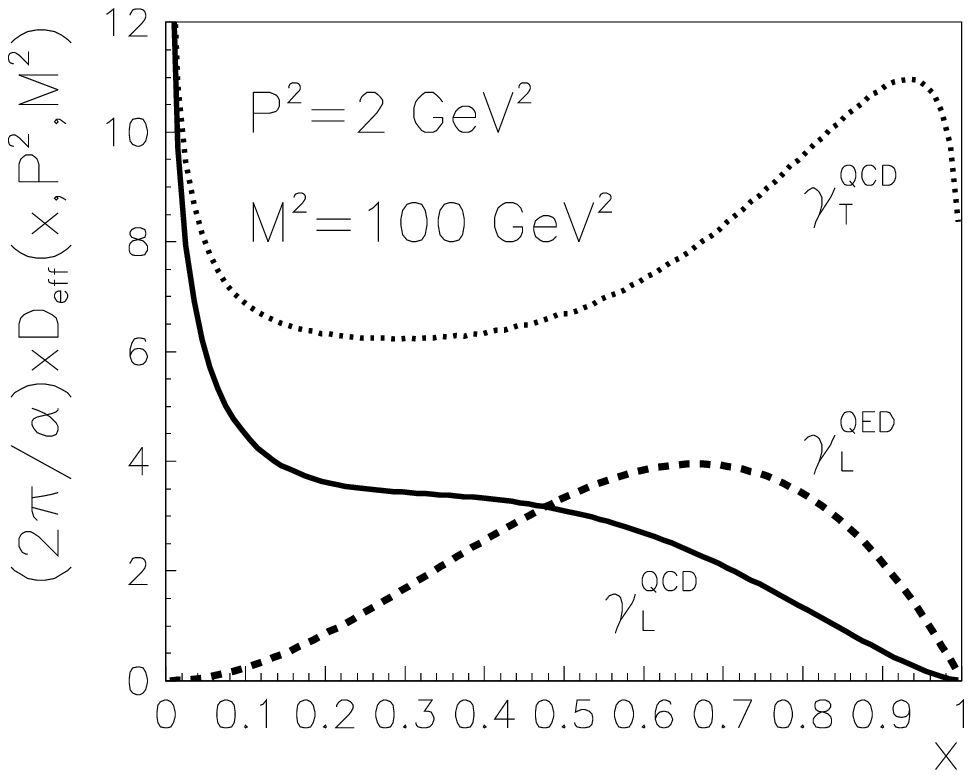,width=7cm}}
\put(75,0){\epsfig{file=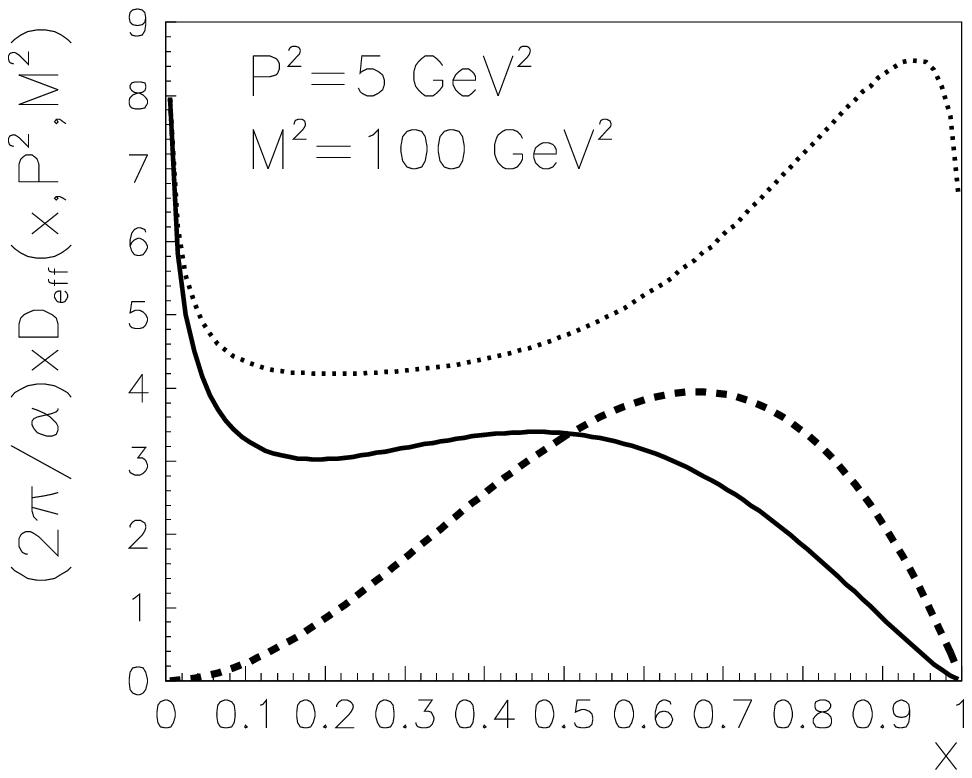,width=7cm}}
\end{picture}
\caption{Comparison of the contributions of resolved $\gamma_T^*$ and
$\gamma_L^*$ to $D_{\mathrm{eff}}$ defined in (\ref{deff}) for
$M^2=100$ GeV$^2$ and
$P^2=2,5$ GeV$^2$. QED and QCD formulae discussed in the text were
used for $\gamma_L^*$ and SaS1D parameterization for $\gamma_T^*$.}
\label{eff2}
\end{figure}
The measurement of dijet cross sections in ep (and e$^+$e$^-$) collisions
offers currently the best way of investigating interactions of virtual
photons \cite{H1eff,phd}. In general the corresponding cross sections
are given as sums of contributions of all possible parton level
subprocess. The simplest way of demonstrating the importance of the
contributions of resolved $\gamma_L^*$ employs the approximation
\cite{Chris} in which dijet cross sections are expressed in terms of
single {\em effective parton distribution function} of the photon
($\gamma_T^*$ or $\gamma_L^*$)
\begin{equation}
D_{\mathrm{eff}}(x,P^2,M^2)\equiv
\sum_{i=1}^{n_f}\left(q_i(x,P^2,M^2)+
\overline{q}_i(x,P^2,M^2)\right)+\frac{9}{4}G(x,P^2,M^2),
\label{deff}
\end{equation}
where the factorization scale $M$ is conventionally identified with
(a multiple of) jet $E_T$: $M=\kappa E_T$.
In Fig. \ref{eff2} the contributions to $D_{\mathrm{eff}}$ of
$\gamma_L^*$, evaluated with both QED and QCD formulae for its PDF,
are compared to those of $\gamma_T^*$ using SaS1D parameterization
\cite{sas1}. The comparison is performed for two pairs of $P^2$ and
$M^2$ typical for HERA experiments. In addition to softening effects at
large $x$, QCD improved PDF of $\gamma_L^*$ give sizable contribution
to $D_{\mathrm{eff}}$ at small $x$ that comes from the gluon content
of $\gamma_L^*$. Fig. \ref{eff2} moreover suggests that in the region
accessible at HERA the contributions of resolved $\gamma_L^*$ are
numerically important, particularly after incorporating QCD effects
in its PDF.

After this simple but approximate estimate of the contributions
of resolved $\gamma_L^*$, I now turn to the evaluation of dijet cross
sections at HERA using complete LO QCD formalism as implemented in
HERWIG 5.9 event generator. To include the effects of resolved
$\gamma_L^*$ I have added the option of generating the flux of
$\gamma_L^*$ combined with the call to QED or QCD improved PDF of
$\gamma_L^*$. For $\gamma_T^*$ the SaS1D PDF \cite{sas1} were used.
\begin{figure*}[t]
\epsfig{file=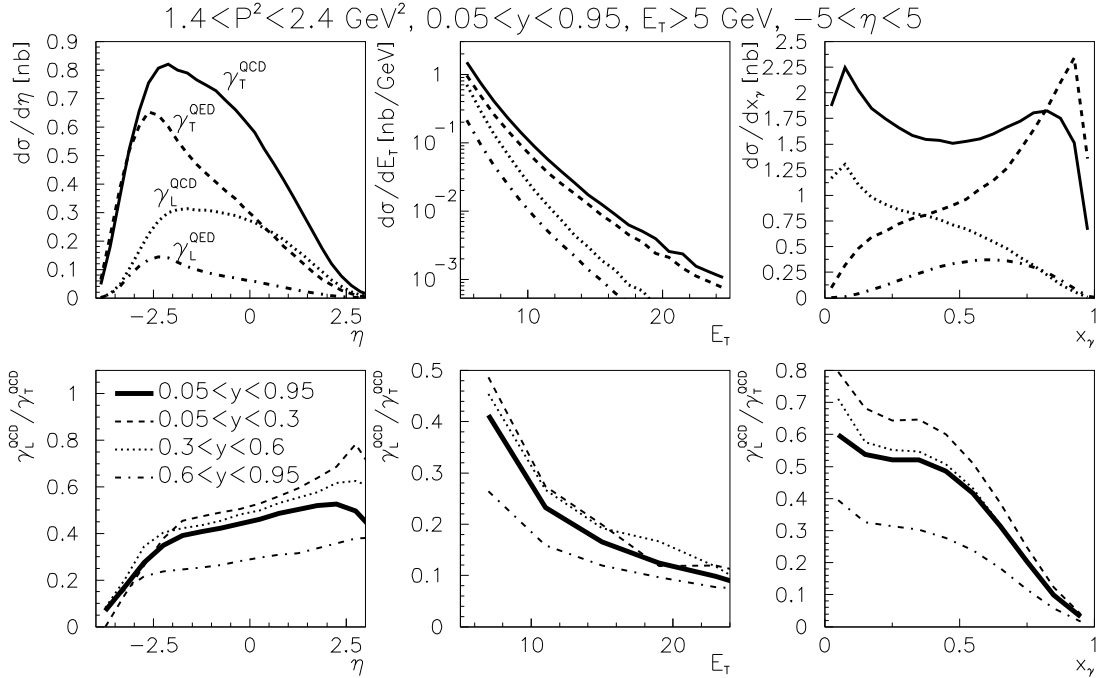,width=\textwidth}
\caption{Upper three plots: dijet cross sections, corresponding to
resolved $\gamma_T^*$ and $\gamma_L^*$ plotted as functions of
$\eta,E_T$ and $x_{\gamma}$ for
$1.4\le P^2\le 2.4$ GeV$^2$, $0.05\le y\le 0.95$, $E_T\ge 5$ GeV,
without any restriction on $\eta$.
Lower three plots: the corresponding ratia of the contributions of
$\gamma^*_L$ and $\gamma_T^*$, integrated over the whole interval
$0.05\le y\le 0.95$, as well as in three indicated subintervals.}
\label{qcd11}
\end{figure*}
\begin{figure*}[t]
\epsfig{file=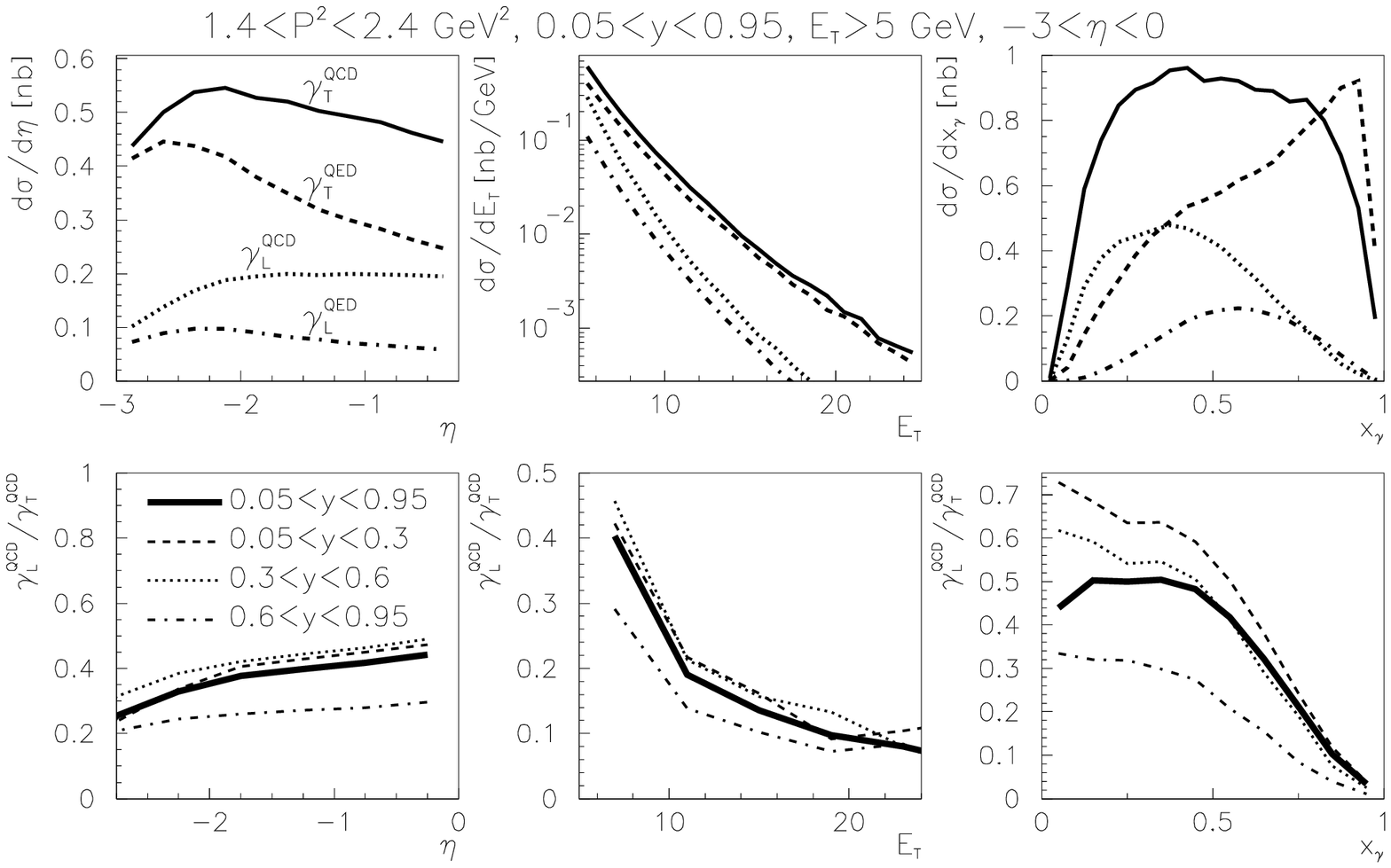,width=\textwidth}
\caption{The same as in Fig. \ref{qcd11}, but the restricted region
$-3\le\eta\le 0$.}
\label{limqcd11}
\end{figure*}
The dijet cross sections were evaluated for
$0.05\le y\le 0.95$ in three windows of $P^2$
$$1.4\le P^2\le 2.4~{\mathrm {GeV}}^2,~
2.4\le P^2\le 4.4~{\mathrm {GeV}}^2,~
4.4\le P^2\le 10~{\mathrm{GeV}}^2$$
and imposing the following cuts on jet $E_T$ (all quantities are in
$\gamma^*$p cms)
$$E_T^{(1)},E_T^{(2)}\ge E_T^c,
~E_T^c=5,10~{\mathrm{GeV}}.$$
The effects of H1 and ZEUS detector acceptances were taken into account
by performing all calculations without any restrictions on $\eta$ as well
for $-3\le \eta\le 0$.

The results presented in Figs. \ref{qcd11} and \ref{limqcd11}
correspond to parton level calculations in the first window of $P^2$,
without and with the mentioned cuts on $\eta$. The
characteristic dependence of the contributions of resolved
$\gamma_L^*$ on $y$ is illustrated by plotting for each of the
distributions in $\eta,E_T$ and $x_{\gamma}$ also its ratio to
that of $\gamma_T^*$ for the whole interval $0.05\le y\le 0.95$, as
well as for three indicated subintervals. Except for $x_{\gamma}$
close to $1$, QCD improved PDF of $\gamma_L^*$ enhance its
contributions to dijet cross sections compared to those based on
the purely QED. For $y\lesssim 0.5$ they amount to about $50\%$ of
those of $\gamma_T^*$. For $x_{\gamma}\lesssim 0.2$ this number
increases further up to about $70\%$. Reducing the range of $\eta$
to $-3\le\eta\le 0$ affects
mainly the distribution ${\mathrm{d}}\sigma/{\mathrm{d}}x_{\gamma}$
by suppressing it at both edges of the phase space. The ratia of the
contributions of $\gamma_L^*$ and $\gamma_T^*$ are, however, affected
only little by this cut.

Increasing the photon virtuality $P^2$ enhances, approximately uniformly
in the whole phase space, the relative importance of $\gamma_L^*$ with
respect to $\gamma_T^*$. On the contrary, rising the threshold $E_T^c$
from $5$ GeV to $10$ GeV reduces it by a factor of about 2, since large
$E_T$ require large $x_{\gamma}$, where quarks from $\gamma_T^*$ dominate.

The effects of hadronization on parton level results discussed above
have been studied in detail in \cite{phd}. They are reasonably small
($\lesssim 10-20$\%) and model independent in the region
$-2.5\lesssim \eta$ but turn large and model dependent below
that value. For the comparison with theoretical calculations
the lower limit on accessible range of $\eta$ enforced by H1 and ZEUS
acceptances presents therefore no real restriction. On the other hand,
it would be very useful to push the upper limit on $\eta$ above
$\eta\simeq 0$ since the relevance of $\gamma_L^*$ grows with $\eta$.

Summarizing the message of this Section, we conclude
that for $\Lambda^2\ll P^2\ll E_T^2$:
\begin{itemize}
\item The contributions of $\gamma_L^*$ are substantial, particularly
 for small $y$, large $P^2$, low $E_T$ and small $x_{\gamma}$.
\item
The cuts enforced by H1 and ZEUS acceptances reduce the sensitivity
to $\gamma_L^*$, but its contributions still make up typically
$50\%$ of those of $\gamma_T^*$ and can be identified by
their characteristic $y$ and $P^2$ dependencies.
\end{itemize}

\section*{Conclusions}
The contributions of resolved $\gamma_L^*$ to dijet production in ep
collisions at HERA were evaluated using QCD improved PDF of $\gamma_L^*$
constructed recently. In the region accessible at HERA they turn out to
sizable, amounting typically to $50$\% of those from $\gamma_T^*$,
and depend sensitively of $y,E_T$ and $x_{\gamma}$.

\vspace*{0.3cm}
\noindent
Work performed under the project LN00A006 of the Ministry of Education of
the Czech Republic

\end{document}